# Negative dispersion medium at 1064 nm in an optomechanical resonator for enhancing the sensitivity bandwidth in a gravitational-wave detector

**Minchuan Zhou,[1] and Selim M. Shahriar[1,2,*]**

[1]*Department of Physics and Astronomy, Northwestern University, Evanston, IL 60208, USA*

[2]*Department of EECS, Northwestern University, Evanston, IL 60208, USA*

*Email: shahriar@northwestern.edu

**Abstract** Recently, we had proposed an optically-pumped five-level Gain EIT (GEIT) system, which has a transparency dip superimposed on a gain profile and exhibits a negative dispersion suitable for the white light cavity (WLC) enhanced interferometric gravitational wave detector [Phys. Rev. D. **92**, 082002 (2015)]. Using this system as the negative dispersion medium (NDM) in the WLC-SR (signal recycling) scheme, we get an enhancement in the quantum noise (QN) limited sensitivity-bandwidth product by a factor of ~18. We have also shown how to realize such a system in practice using Zeeman sublevels in [87]Rb at 795nm [Opt. Commun. 402, 382-388 (2017)]. However, aLIGO operates at 1064nm and suitable transitions in Rb or other alkali atoms are not available at this wavelength. Therefore, it is necessary to consider a system that is consistent with the operating wavelength of aLIGO. Here, we present the realization of such an NDM at 1064nm with a microresonator, which supports optomechanical interaction. A strong control field is applied at a higher frequency, and, under certain conditions, a probe field at a lower frequency experiences a peak at the center of an absorption profile, and a negative dispersion in the transmission. Unlike in the GEIT case, we use the compound-cavity signal-recycling (CC-SR) scheme, where an auxiliary mirror is inserted in the dark port of the detector, and show that the enhancement factor can be as high as ~15. However, using the parameters required for the sensitivity enhancement, the optomechanical system enters an instability region where the control field is depleted. We present an observer based feedback control process used to stabilize the CC-SR system.

## I. INTRODUCTION

Previously, we had proposed an interferometric gravitational wave (GW) detector using a white light cavity [1,2,3,4,5,6,7,8] for signal recycling in the advanced Laser Interferometric Gravitational-Wave Observatory (aLIGO). The key element in the WLC is a negative dispersion medium (NDM), used to compensate the phase variation due to change in frequency in the arm cavities, including optomechanical effects. One way to realize such an NDM makes use of non-degenerate Zeeman sublevels in cold [87]Rb atoms [9]. The resulting susceptibility shows a transparency dip on top of a gain profile (GEIT) and a negative dispersion suitable for the phase compensation.

However, application of the [87]Rb-based GEIT system to aLIGO has several drawbacks. First, the [87]Rb GEIT system operates at 795nm, different from the operating wavelength of current LIGO at 1064nm. The prospect of a future GW detector operating at 795nm is, to the best of our knowledge, not being envisioned by anyone. Second, the Rb-GEIT process requires cold atoms and a rather large density-length product of $1.25 \times 10^{18} \text{m}^{-2}$, which is a challenging requirement to meet experimentally.

In this paper, we describe a realization of the NDM that operates at 1064nm using an optomechanical resonator. In Sec. II, we describe and theoretically model an optomechanical-resonator-based system that produces a negative dispersion at 1064nm, while adding minimal noise, similar to the case for GEIT. In Sec. III, we analyze the quantum noise (QN) of the optomechanical system, and calculate the QN limited sensitivity of the aLIGO apparatus incorporating this system as the NDM in the signal recycling cavity using the Langevin noise operator model. In Sec. IV, we discuss the stabilization of the system. In Sec. V, we summarize the results. In Appendix A, we discuss some details of the observer based feedback control. In Appendix B, we describe a triangular optical cavity that can be used as the beam splitter/combiner for incorporating the optomechancial system in the CC-SR scheme.

## II. NDM USING THE OPTOMECHANICAL RESONATOR

We introduce below an optomechanical system, in which two optical modes couple to a mechanical mode. As shown in Fig. 1, a microsphere resonator that supports optomechanical interaction is coupled to a tapered optical fiber. The resonator hosts two optical modes $\hat{a}_1$ and $\hat{a}_2$ at frequencies $\omega_{C1}$ and $\omega_{C2}$, respectively. The two optical modes are coupled to an acoustic phonon mode $\hat{b}$ of frequency $\omega_m = \omega_{C2} - \omega_{C1}$, mediated by Brillouin scattering [10,11,12]. A probe field $\hat{s}_{1,IN}$ at the frequency of $\omega_1$ and a control field $\hat{s}_{2,IN}$ at the frequency of $\omega_2$ are sent into the tapered fiber coupler. The probe field excites the lower energy optical mode $\hat{a}_1$, while the strong control field excites the higher energy optical mode $\hat{a}_2$. The outputs from the waveguides are $\hat{s}_{1,OUT}$ and $\hat{s}_{2,OUT}$. As we will see later, the amplitude of the control field determines the strength of the optomechanical coupling, and the probe field experiences an absorption with a narrow transparency window in the transmission profile.

Previously, the Brillouin scattering induced transparency (BSIT) has been observed experimentally in the optomechanical system similar to the configuration shown in Fig. 1, where the control field excites the lower energy optical mode and the probe field excites the higher energy optical mode [12]. In contrast with the slow light effect in BSIT, we here observe a negative dispersion and a fast light effect. In analogy to the physical intuition of BSIT [12], such a transparency can be described qualitatively as follows. Let us denote by $n_p$ and $n_m$ the number of probe photons and the number of phonons, respectively, as indicated schematically in Fig. 1 (b). Consider now two photon-phonon dual Fock states, one represented as $|n_p, n_m\rangle$ and the other as $|n_p, n_m - 1\rangle$. The coupling of a probe photon into the resonator leads to a transition from $|n_p, n_m\rangle$ to $|n_p + 1, n_m\rangle$. On the other hand, the Stokes scattering of the control field results in the transition from $|n_p, n_m - 1\rangle$ to $|n_p + 1, n_m\rangle$. We can define the dark state and the bright state as $|D\rangle \equiv [\Omega_1 |n_p, n_m - 1\rangle - \Omega_2 |n_p, n_m\rangle] / \sqrt{\Omega_1^2 + \Omega_2^2}$ and $|B\rangle \equiv [\Omega_2 |n_p, n_m - 1\rangle + \Omega_1 |n_p, n_m\rangle] / \sqrt{\Omega_1^2 + \Omega_2^2}$, respectively, where $\Omega_1$ ($\Omega_2$) is the rate at which $|n_p, n_m\rangle$ ($|n_p, n_m - 1\rangle$) couples to $|n_p + 1, n_m\rangle$. The bright state couples strongly to $|n_p + 1, n_m\rangle$, which then decays back to both bright and dark states. On the other hand, the dark state is fully decoupled from $|n_p + 1, n_m\rangle$. This process leads to a steady-state situation where only the dark state is populated, thus producing the transparency. We emphasize that this interpretation is only qualitative, since a complete interpretation of the transparency in terms of these dual Fock states requires a Monte-Carlo simulation, due to the large number of such states involved. Such a simulation is beyond the scope of this paper, and will be carried out in the future. However, we note that the analysis shown below, using Heisenberg operators for the photons and the phonons inside the resonator, produces a transparency for the probe, consistent with this qualitative interpretation.

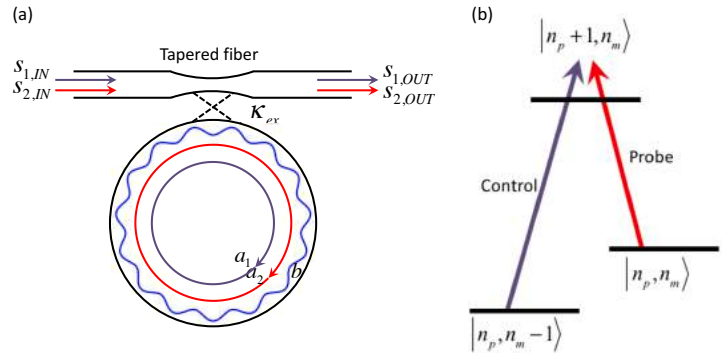

FIG. 1. Schematic illustration of (a) the negative dispersion medium using an optomechanical resonator and (b) Brillouin scattering analogue of Electromagnetically Induced Transparency in a three-level lambda system, in terms of the photon-phonon dual Fock states.

The optomechanical Hamiltonian is given by [13]

$$\hat{H} = \hbar\omega_{C1}\hat{a}_1^\dagger\hat{a}_1 + \hbar\omega_{C2}\hat{a}_2^\dagger\hat{a}_2 + \hbar\omega_m\hat{b}^\dagger\hat{b} + \hat{H}_{INT1} + \hat{H}_{INT2}, \quad (1)$$

where $\hat{H}_{INT1}$ represents the interaction amongst the two optical modes inside the resonator and the phonon mode, while $\hat{H}_{INT2}$ represents the coupling between the fields propagating through the fiber and the optical modes inside the resonator.

$$\hat{H}_{INT1} = -\hbar(\beta\hat{a}_2^\dagger\hat{a}_1\hat{b} + \beta^*\hat{b}^\dagger\hat{a}_1^\dagger\hat{a}_2), \quad (2)$$

$$\hat{H}_{INT2} = i\hbar\sqrt{\kappa_{EX}}(\hat{a}_1^\dagger\hat{s}_{1,IN}e^{-i\omega_1 t} + \hat{a}_2^\dagger\hat{s}_{2,IN}e^{-i\omega_2 t}) + h.c., \quad (3)$$

where $\beta$ is the optomechanical coupling rate, $\kappa_{ex}$ is the waveguide-resonator coupling rate, and the time dependence $e^{-i\omega_1 t}$ and $e^{-i\omega_2 t}$ are separated out from $\hat{s}_{1,IN}$ and $\hat{s}_{2,IN}$. A unitary transformation $\hat{U} = \exp[i\omega_1\hat{a}_1^\dagger\hat{a}_1 t + i\omega_2\hat{a}_2^\dagger\hat{a}_2 t + i(\omega_2 - \omega_1)\hat{b}^\dagger\hat{b}t]$ makes the interaction Hamiltonians time independent, and generates the new Hamiltonian $\hat{H}' = \hat{U}\hat{H}\hat{U}^\dagger - i\hbar\hat{U}\partial\hat{U}^\dagger/\partial t$, written as

$$\hat{H}' = \hbar\Delta_1\hat{a}_1^\dagger\hat{a}_1 + \hbar\Delta_1\hat{a}_2^\dagger\hat{a}_2 + \hbar\Delta_m\hat{b}^\dagger\hat{b} + \hat{H}'_{INT1} + \hat{H}'_{INT2}, \quad (4)$$

where

$$\Delta_j = \omega_j - \omega_{cj}, j=1,2, \quad \Delta_m = (\omega_2 - \omega_1) - \omega_m, \quad (5)$$

$$\hat{H}'_{INT1} = -\hbar(\beta\hat{a}_2^\dagger\hat{a}_1\hat{b} + \beta^*\hat{b}^\dagger\hat{a}_1^\dagger\hat{a}_2), \quad (6)$$

$$\hat{H}'_{INT2} = i\hbar\sqrt{\kappa_{EX}}(\hat{a}_1^\dagger\hat{s}_{1,IN} + \hat{a}_2^\dagger\hat{s}_{2,IN}) + H.c., \quad (7)$$

Then we obtain the equations of motion for the field amplitudes defined as $a_j = \langle\hat{a}_j\rangle$, $s_{j,IN} = \langle\hat{s}_{j,IN}\rangle$, $j=1,2$ and $b = \langle\hat{b}\rangle$ as follows:

$$\dot{a}_1 = -\gamma_1 a_1 - i\beta^* a_2 b^* + \sqrt{\kappa_{EX}} s_{1,IN}, \quad (8)$$

$$\dot{a}_2 = -\gamma_2 a_2 - i\beta a_1 b + \sqrt{\kappa_{EX}} s_{2,IN}, \quad (9)$$

$$\dot{b} = -\gamma_m b - i\beta^* a_1^* a_2, \quad (10)$$

where $\gamma_j = \kappa_j/2 - i\Delta_j, j=1,2$, $\kappa_j = \kappa_{0j} + \kappa_{EX}$ is the damping rate for the optical mode $a_j$, and $\gamma_m = \Gamma/2 - i\Delta_m$. Here $\Gamma$ is the damping rate for the mechanical mode.

The input-output relations, which relates the intra-resonator field $a_j$, and the input and output fields are

$$s_{j,OUT} = s_{j,IN} - \sqrt{\kappa_{EX}} a_j, \text{for } j = 1,2. \tag{11}$$

Since the control field is assumed to be strong and the dissipation caused by the optomechanical coupling is very small in comparison, we neglect the second term in Eq. (9). Solving Eqs. (8)-(10) in steady state, we get

$$a_2 = \frac{\sqrt{\kappa_{EX}}}{\gamma_2} s_{2,IN}, b = \frac{-i\beta^* a_1^* a_2}{\gamma_m}, a_1 = \frac{\sqrt{\kappa_{EX}}}{\gamma_1 - \alpha} s_{1,IN}, \tag{12}$$

where

$$\alpha = \frac{|g|^2}{\gamma_m^*}, \quad g = \beta a_2^*. \tag{13}$$

The effective coupling $g$ can be governed by the control field $a_2$. Using Eq. (11), we get the transmission of the probe field,

$$t_1 \equiv |t_1| e^{i\theta_1} = \frac{s_{1,OUT}}{s_{1,IN}} = 1 - \frac{\kappa_{EX}}{\gamma_1 - \alpha}. \tag{14}$$

We consider the case where the control field $s_{2,IN}$ is parked on the resonator mode $a_2$ with fixed detuning $\Delta_2 = 0$, i.e. $\omega_2 = \omega_{C2}$. The wavelength of the optical mode $a_1$ is chosen to be $\lambda = 1064 nm$, corresponding to the operating wavelength of LIGO, so that $\omega_{C1} = 2\pi c / \lambda$. The frequency of the probe field is scanned around $\omega_{C1}$, with detuning $\Delta_1 = \omega_1 - \omega_{C1}$. We consider the case where $(\omega_{C2} - \omega_{C1})/(2\pi) = \omega_m /(2\pi) = 1 GHz$.

When the phonon decay rate is large $\Gamma/(2\pi) = 1 GHz$, we see a gain peak in the transmission of the probe (Fig. 2), corresponding to Brillouin gain, and a normal dispersion in the phase response $\theta_1$. With a much smaller phonon decay rate, $\Gamma = 1 Hz$, we observe an EIT-like transmission with a negative dispersion (Fig. 3) [14]. With the choice of parameters $\omega_{C2} - \omega_{C1} = \omega_m, \kappa_0 = \kappa_1 - \kappa_{EX} = 1 Hz, \kappa_{EX} = 0.13 MHz, g = 30 kHz$, we plot the results for the transmission and phase response in Fig. 3.

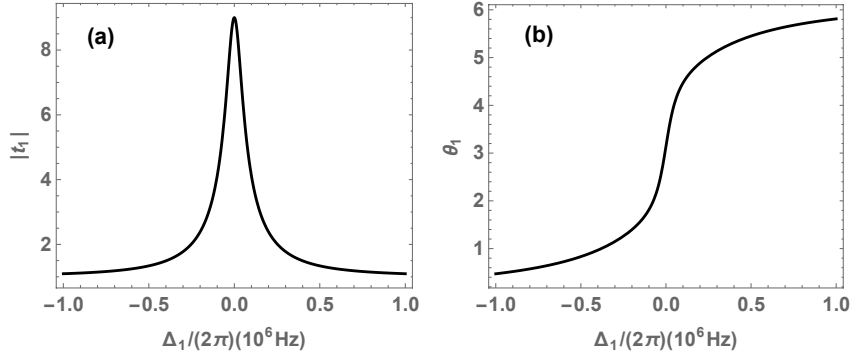

FIG. 2. (a) The transmissivity $|t_1|$ and (b) phase shift $Arg[t_1] = \theta_1$ as a function of detuning $\Delta_1 = \omega_1 - \omega_{C1}$ in the optomechanical resonator when $\kappa_0 = \kappa_1 - \kappa_{EX} = 0.5 MHz$, $\kappa_{EX} = 0.5 MHz$, $\Gamma = 1GHz$, and $g = 15MHz$.

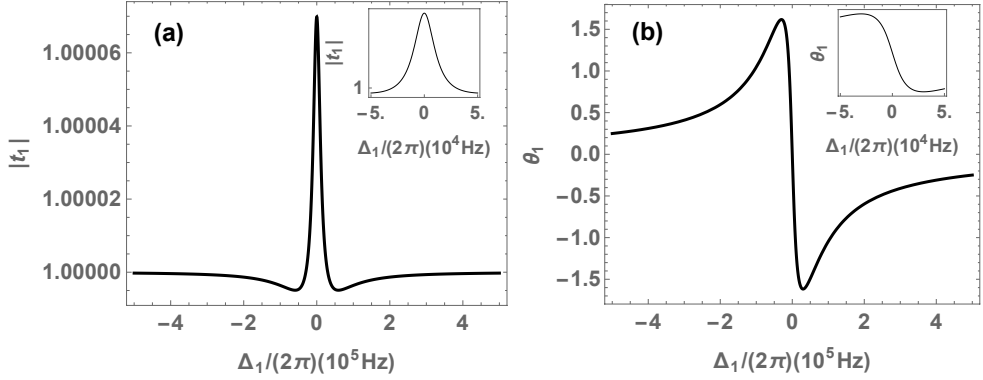

FIG. 3. (a) The transmissivity $|t_1|$ and (b) phase shift $\theta_1$ as a function of detuning $\Delta_1 = \omega_1 - \omega_{C1}$ in the optomechanical resonator when $\kappa_0 = 1Hz$, $\kappa_{EX} = 0.13MHz$, $\Gamma = 1Hz$, and $g = 30kHz$. The insets show the plots on a smaller scale.

## III.  CALCULATING THE QUANTUM NOISE

The optomechanical system we considered in the previous sections shows a transparency window in the center of an absorption profile. In order to analyze the QN from this system, we use the Langevin noise operator model. The quantum Langevin equations are written as follows:

$$\dot{\hat{a}}_1 = -\left(\frac{\kappa_1}{2} - i\Delta_1\right)\hat{a}_1 - i\beta^*\hat{a}_2\hat{b}^\dagger + \sqrt{\kappa_{EX}}\hat{s}_{1,IN} + \sqrt{\kappa_0}\hat{f}_1, \tag{15}$$

$$\dot{\hat{a}}_2 = -\left(\frac{\kappa_2}{2} - i\Delta_2\right)\hat{a}_2 - i\beta\hat{a}_1\hat{b} + \sqrt{\kappa_{EX}}\hat{s}_{2,IN} + \sqrt{\kappa_0}\hat{f}_2, \tag{16}$$

$$\dot{\hat{b}}^\dagger = -\left(\frac{\Gamma}{2} + i\Delta_m\right)\hat{b}^\dagger + i\beta\hat{a}_1\hat{a}_2^\dagger + \sqrt{\Gamma}\hat{f}_b^\dagger, \tag{17}$$

where $\hat{f}_\beta (\beta = 1,2,b)$ are the Langevin noise operators [15] that satisfy the following relation:

$$\left\langle \hat{f}_\beta(t)\hat{f}_\beta^\dagger(t')\right\rangle = (\bar{n}_\beta + 1)\delta(t-t'), \quad \left\langle \hat{f}_\beta^\dagger(t)\hat{f}_\beta(t')\right\rangle = \bar{n}_\beta \delta(t-t'), \tag{18}$$

Here $\bar{n}_\beta$ is the occupation number of the thermal reservoir. Since the frequencies of the optical modes are very high, compared to the thermal frequency (defined as $k_B T/\hbar$, where $k_B$ is the Boltzmann constant, and $T$ is the temperature) even at room temperature, the corresponding occupation numbers $\bar{n}_1$ and $\bar{n}_2$ can be assumed to be zero. In general, the phonon mode has non-zero thermal occupation $\bar{n}_b \neq 0$.

The following conventions are used for Fourier transforms:

$$\hat{f}_\beta(\omega) = \int_{-\infty}^{\infty} e^{i\omega t}\hat{f}_\beta(t)dt, \beta = 1,2,b. \tag{19}$$

We can then solve the set of equations in the frequency domain, which gives

$$b^\dagger = \frac{iga_1 + \sqrt{\Gamma}\hat{f}_b^\dagger}{\gamma_m^*}, \tag{20}$$

$$\hat{a}_1(\omega) = \frac{\sqrt{\kappa_{EX}}\hat{s}_{1,IN}(\omega)}{\gamma_1 - |g|^2/\gamma_m^*} + \frac{\sqrt{\kappa_{01}}\hat{f}_1(\omega)}{\gamma_1 - |g|^2/\gamma_m^*} + \frac{-ig\sqrt{\Gamma}\hat{f}_b^\dagger}{\gamma_1\gamma_m^* - |g|^2}, \tag{21}$$

where $\gamma_m^* = \Gamma/2 + i(\Delta_m - \omega)$ and $\gamma_1 = \kappa_1/2 - i(\Delta_1 + \omega)$. Using the input-output relation that

$$\hat{s}_{1,OUT}(\omega) = \hat{s}_{1,IN}(\omega) - \sqrt{\kappa_{EX}}\hat{a}_1(\omega), \tag{22}$$

we get that:

$$\hat{s}_{1,OUT}(\omega) = C_1\hat{s}_{1,IN}(\omega) + C_2\hat{f}_1(\omega) + C_3\hat{f}_b^\dagger(\omega) \tag{23}$$

where

$$C_1 = \left(1 - \frac{\kappa_{EX}}{\gamma_1 - |g|^2/\gamma_m^*}\right), \quad C_2 = -\frac{\sqrt{\kappa_{01}\kappa_{EX}}}{\gamma_1 - |g|^2/\gamma_m^*}, \quad C_3 = \frac{ig\sqrt{\Gamma\kappa_{EX}}}{\gamma_1\gamma_m^* - |g|^2}, \tag{24}$$

The spectral density $S_{1,OUT}(\omega)$ of the output field can be derived using

$$\left\langle \hat{s}_{1,OUT}(\omega)\hat{s}_{1,OUT}^\dagger(\omega') + \hat{s}_{1,OUT}^\dagger(\omega)\hat{s}_{1,OUT}(\omega')\right\rangle \equiv 2\pi\delta(\omega-\omega')S_{1,OUT}(\omega), \tag{25}$$

which gives

$$S_{1,OUT}(\omega)=|C_1|^2 S_{1,IN}(\omega)+Q_{ADD}, Q_{ADD}=|C_2|^2+|C_3|^2(2\bar{n}_b+1), \tag{26}$$

where $|C_1|$ is essentially the same as the transmissivity $|t_1|$ that we calculated in Eq. (14), and $Q_{ADD}$ represents the amount of noise due to the interaction with the pump and the resonator. We assume a temperature of 30mK for the thermal reservoir for the phonon mode, which can be achieved using a Helium dilution refrigerator. The spectral shape of $Q_{ADD}$ using the parameters of the optomechanical system same as those in Fig. 3 is shown in Fig. 4.

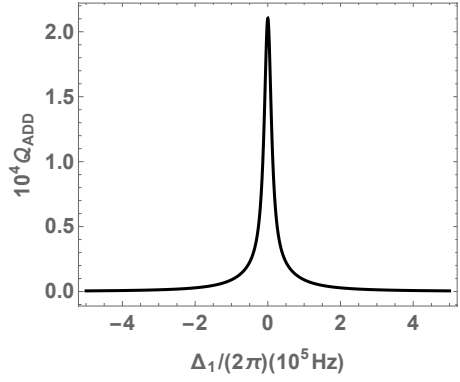

FIG. 4. Plot of the noise $Q_{add}$ as a function of the frequency $\Delta_1$.

The compound-cavity signal-recycling (CC-SR) [1] interferometric gravitational wave detector is schematically shown in Fig. 5, where the reflectivity of the SR mirror ($M_{SR}$) is matched to that of the input test mass mirror ($M_1$) and the length of the SRC is chosen so that the SRC effectively disappears in the advanced Laser Interferometric Gravitational-Wave Observatory (aLIGO) [16]. We then add an auxiliary mirror ($M_{AUX}$). In order to achieve an enhancement in the quantum noise (QN) limited sensitivity-bandwidth product, we need to use an NDM with a dispersion compensating the phase variation as the frequency changes, and with vanishingly small QN around the center of the dispersion [8].

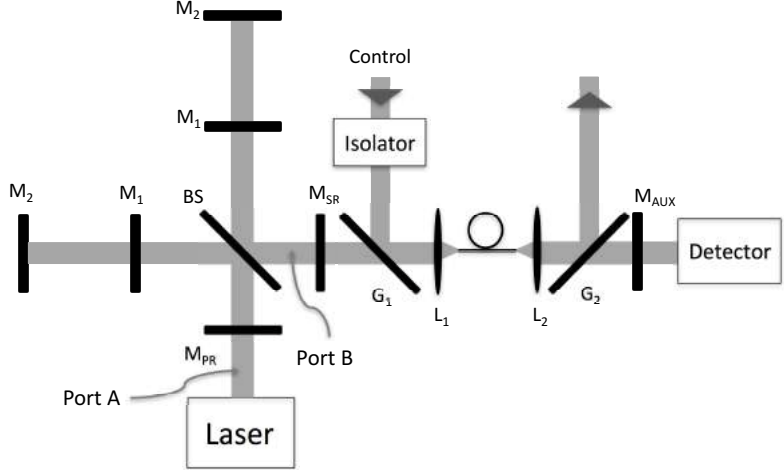

FIG. 5. The CC-SR design using the optomechanical system as the NDM. A negative dispersive medium realized using an optomechanical ring resonator is inserted in the auxiliary cavity formed by $M_{AUX}$ and $M_{SR}$. The control field is sent to an isolator and then combines with the output from BS by a beam splitter/combiner ($G_1$). The resulting field is coupled to the tapered fiber by a lens. The output from the tapered fiber is expanded using another lens. Using another a beam splitter/combiner ($G_2$), the control field is filtered out. If we use another control field in the backward direction (from the detector to BS), we can send the field in through $G_2$, which is then filtered out by $G_1$.

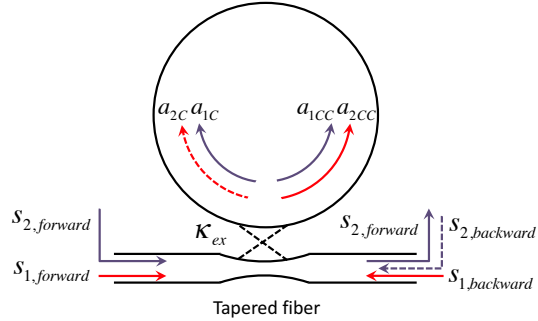

FIG. 6. Schematic illustration of the fields in the optomechanical system: $s_{1,forward}$ ($s_{1,backward}$) is the probe field in the forward (backward) direction, which excites the counter-clockwise (clockwise) field $a_{1CC}$ ($a_{1C}$) in the microresonator. For simplicity, we consider the case that the control field $s_{2,forward}$ is applied only in the forward direction, which excites the counter-clockwise field $a_{2CC}$ in the microresonator. In general, we can also apply a control field $s_{2,backward}$ in the backward direction.

For simplicity, we assume that a control field is applied only in the forward direction (going from the BS to the detector). Therefore the $a_2$ mode is excited only in the counter-clockwise direction (when looking into the diagram). This is illustrated schematically in Fig. 6. Since the probe field excites the clockwise $a_1$ field when propagating backward in the auxiliary cavity, and this clockwise field would not couple to the counter-clockwise $a_2$ field, the field in the interferometer would experience a loss. To make this loss small, we need to use a microresonator that supports very small optical loss, $\kappa_0$. This restraint can be relaxed if an optical circulator that has a very small insertion loss is available. In the following, we assume the parameters to be: $\kappa_0 = 1 Hz$, $\kappa_{EX} = 0.13 MHz$, $\Gamma = 1 Hz$, and $g = 30 kHz$.

To calculate the QN of the CC-SR scheme, we represent the fields as the amplitudes of the two-photon modes following the two-photon formalism developed by Caves and Schumaker [17,18], and derive the input-output relation between the principal noise input and the signal and noise output [8]. In order to calculate the QN due to the absorption and amplification in the NDM, we use the Langevin noise operator model. Using the same method as in Sec. III in Ref. 8, we plot the resulting QN curves of the CC-SR scheme in Fig. 5. The QN curves for the CC-SR scheme (shown as the green curves in Fig. 7) stay well below the SQL line, and show an enhancement in the sensitivity-bandwidth product by a factor of ~15 compared to the curve for the GW detector in the SR configuration with the highest sensitivity result (shown as red dashed curve) predicted by Bunanno and Chen [19].

In practice, in order to implement such an NDM in the current aLIGO, we need to use a lens ($L_1$) to couple the output from the beam splitter (BS) to the tapered fiber in Fig. 1, and then use another lens ($L_2$) to expand the beam. The specific scheme for incorporating the optomechanical system in aLIGO is shown in Fig. 5. The control field is sent through an isolator and then combined with the output from the beam splitter (BS) by a beam splitter/combiner ($G_1$). The resulting field is coupled to the tapered fiber by a lens. The output from the tapered fiber is expanded using another lens. Using another beam splitter/combiner ($G_2$), the control field is filtered out. We assume that the insertion loss due to the additional optical components can be minimal, in the range of 0.02% to 0.05%. For the case where the additional noise is 0.02%, the resulting curves are shown in black in Fig. 7, with the enhancement factor dropping to ~14. For the case where the additional noise is 0.05%, the enhancement factor drops to ~12 (not shown). In Fig. 7(b), we show how the factor of enhancement in the sensitivity-bandwidth product depends on the insertion loss.

For the beam splitter/combiner discussed in Appendix B, assuming the reflectivities of each of the partial reflectors to be 99.999% [20], the signal, in the process of reflection, will experience a loss of 0.001%. For two stages (input and output of the micro-resonator), the total loss would be 0.002%. It is also possible to make lenses with an anti-reflection coating such that the reflection loss on each surface could be as small as 0.005% [21]. Accounting for four lens surfaces, the total loss would be 0.02%. Similar coating can possibly be developed for fiber tips as well. Thus, reflection losses from the input to the taper fiber and its output would contribute to additional reflection loss of 0.01%. Combining all these, we get a reflection loss total of 0.03%. Then comes the question of how well a mode can be coupled in to a fiber (coupling out of fiber can safely be considered to essentially perfect, except for residual reflection loss, which we addressed above). If we want to achieve a net loss no more than 0.05%, then this coupling loss needs to be less than 0.02%, or 0.01% on each side. Producing such a high efficiency coupling into the optical fiber is a difficult, but perhaps not insurmountable, challenge that needs to be overcome in order to implement the scheme proposed here with maximum efficacy.

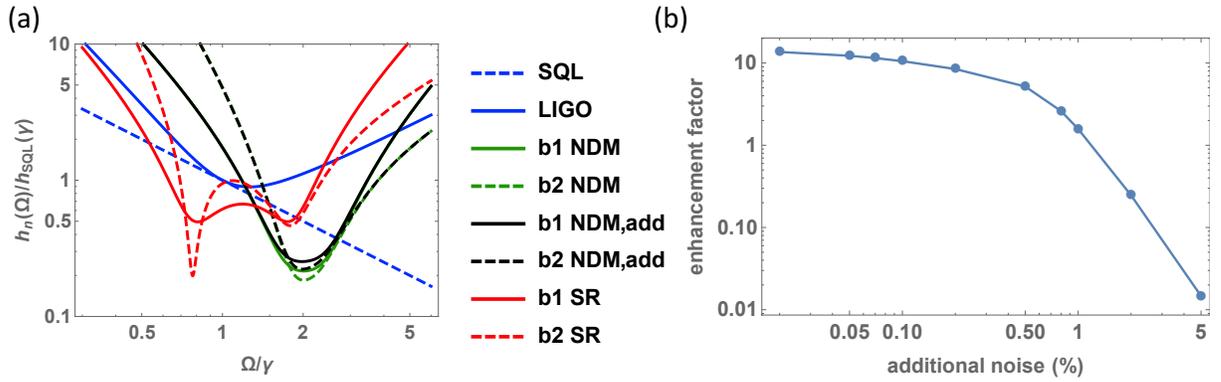

FIG. 7. (a) Log-log plot of the normalized QN $h_n(\Omega)/h_{SQL}(\gamma)$ of the CC-SR scheme versus $\Omega/\gamma$ for the first quadrature b1 and second quadrature b2, following the two-photon formalism developed in Refs. 17 and 18. Here $h_n(\Omega)$ is the square root of the noise spectral density for the GW signal at a sideband frequency $\Omega$, $h_{SQL}(\gamma)$ is the standard quantum limit for GW detection at a sideband frequency $\Omega = \gamma$, where $\gamma$ is the half bandwidth of the arm cavity of the detector. The green curves represent the QN for the CC-SR using the optomechanical system as the NDM. The noise curves with an additional $2\times10^{-4}$ noise (denoted by "NDM, add") are shown in black lines. The red curves represent the QN for the GW detector with SR. The noise curve for LIGO and the standard quantum limit (SQL) curve are plotted in blue. For additional details underlying the notations used here, see, for example, Ref. 8 or Ref. 19. (b) Plot of the enhancement factor in the sensitivity-bandwidth product as a function of the amount of insertion loss.

Another possible concern with the use of the optical fiber in this manner is that the high-order modes in the dark port will not couple well into the fiber and be reflected back to the interferometer. Most of the light scattered from these modes would not resonate in the arm cavities. Still, even a residual amount of such back-scattering is known to introduce an unacceptable amount of noise in the observation band of interest, if the surface causing the back-scattering is not sufficiently vibration-isolated in this band. We envision a scheme wherein a pinhole will be placed in front of the optical fiber, in order to minimize the degree of back-scattering. Furthermore, both the pinhole and the fiber assembly would be isolated from vibration in the observation band to a degree that would make the effect of the residual backscattering insignificant.

Finally, we note that the parameters we have chosen for the micro-resonator are highly demanding, and may be difficult to realize experimentally given the current state of the art. For the mechanical modes, the required quality factor, $Q_m$, is $\sim 10^9$. There are some publications in the literature about the prospect of realizing resonators with such a high quality factor [22,23]. In Ref. 22, the authors show their experimental result for $Q_m \sim 10^8$ at room temperature and notes that this allows one to speculate that coupling new generation of resonator to low temperature baths may yield $Q_m \sim 10^9$. In Ref. 23, the authors show from simulation that a strong optical trap allow a $Q_m$-enhancement factor of ~1500, which opens up the possibility of realizing a $Q_m \sim 10^9$.

The optical quality factor, $Q_{OP}$, we have proposed is $\sim 3 \times 10^{14}$, which is indeed much higher than the best reported to date [24,25]. In both Refs. 24 and 25, $Q_{OP} \sim 10^{10}$ was reported. As noted in Ref. 24, $Q_{OP}$ is limited to this value due to optical loss in the material. For fused silica used in this work, the optical loss at 633 nm is 7 dB/km, dominated by Rayleigh scattering (5 dB/km), and the rest (2 dB/km) attributable to material absorption. The Rayleigh scattering is due to thermal and frozen fluctuations in density. Using crystalline material and operating at extremely low temperatures as proposed in this paper, it may be possible to suppress Rayleigh scattering strongly. Even then, it would be necessary to employ a material which has a much lower material absorption at 1064 nm in order to realize the high value of $Q_{OP}$ proposed here. We are also working on possible variations of the proposed scheme that can possibly relax this constraint significantly.

## IV. STABILIZATION OF THE SYSTEM

When we analyze the optomechanical system using Eqs. (8)-(9), we use the non-depletion approximation for the control field. Under this approximation, we would have the steady state solution:

$$a_1 = -ig^* b^* / \gamma_1 + \sqrt{\kappa_{EX}} s_{1,in} / \gamma_1, \qquad (27)$$

The equation of motion for the phonon mode b is then:

$$\dot{b} = -(\gamma_m - |g|^2 / \gamma_1^*) b - ig^* \sqrt{\kappa_{EX}} s_{1,in} / \gamma_1^*, \qquad (28)$$

With the parameters we used in the results in the previous sections, we would have $\gamma_m \ll |g|^2 / \gamma_1^*$, which causes mechanical instability in the system [13, 26]. In order to stabilize the system [27], we consider the equations of

motion for the probe mode $a_1$ and the phonon mode $b$ in the optomechanical system, and the gravitational wave sideband mode $d$ [8]. Since the mirrors M$_1$ and M$_{SR}$ effectively disappear for the relevant range of frequencies, the main interferometer [defined as the system shown in Fig. 5 without the micro-resonator and the optical elements G$_1$, G$_2$, L$_1$, and L$_2$] can be mapped into a two-mirror cavity [28]. The Hamiltonian for the main interferometer is [26, 28]

$$\hat{H}_{MAIN} = \hbar(\omega_0 + \Delta_{MAIN})\hat{d}^\dagger\hat{d} + \frac{\hat{P}^2}{2M} - \hbar G_0(\hat{d}^\dagger + \hat{d})\hat{X} + \hat{X}F_{GW} + \hat{H}_{\gamma_{MAIN}} + \hat{H}_{\gamma_{INS}}, \tag{29}$$

Here $\omega_0$ is the carrier frequency, $\Delta_{MAIN}$ is detuning of the mode $d$ away from the carrier, and $\hat{X}$ and $\hat{P}$ are the position and momentum operators, respectively, for the differential motion of the mirrors. $F_{GW}(t) = ML\ddot{h}(t)/2$ is the GW force, where $M$ is the mass of M$_1$ and M$_2$. The parameter $G_0$ is the main interferometer optomechanical coupling rate defined as $G_0 \equiv \omega_0 \bar{d}/L$, where $\bar{d} = [2P_{arm}L/(\hbar\omega_0 c)]^{1/2}$ (which represents the average number of photons in the arms), with $P_{arm}$ being the circulating power of the carrier light inside each arm, and L being the length of each arm. $\hat{H}_{\gamma_{MAIN}}$ accounts for the loss in the main interferometer, and $\hat{H}_{\gamma_{INS}}$ accounts for the insertion loss necessary for adding the micro-resonator. The coupling between the field $d$ in the main interferometer and the field $a_1$ in the micro-resonator can be described by the Hamiltonian:

$$\hat{H}_{INT} = \hbar\omega_s(\hat{d}^\dagger\hat{a} + \hat{d}\hat{a}^\dagger), \tag{30}$$

Here $\omega_s$ is the coupling rate [26,29], defined as $\omega_s \equiv \sqrt{c\kappa_{EX}/2L}$. As described in Sec. III, we assume the control field only in the forward direction; therefore the counter-clockwise field $a_{1CC}$, the $a_2$ field, and the phonon mode in the micro-resonator are coupled, while the clockwise field $a_{1C}$ is not coupled to $a_2$ and $b$. The resulting equations of motion for the system are:

$$\dot{\hat{a}}_{1CC} = -i\omega_s\hat{d} - \frac{\kappa_0}{2}\hat{a}_{1CC} - ig^*\hat{b}^\dagger + \sqrt{\kappa_0}\hat{a}_{1CC}^{th}, \tag{31}$$

$$\dot{\hat{a}}_{1CC}^\dagger = i\omega_s\hat{d}^\dagger - \frac{\kappa_0}{2}\hat{a}_{1CC}^\dagger + ig\hat{b} + \sqrt{\kappa_0}\hat{a}_{1CC}^{th\dagger}, \tag{32}$$

$$\dot{\hat{a}}_{1C} = -i\omega_s\hat{d} - \frac{\kappa_0}{2}\hat{a}_{1C} + \sqrt{\kappa_0}\hat{a}_{1C}^{th}, \tag{33}$$

$$\dot{\hat{a}}_{1C}^\dagger = i\omega_s\hat{d}^\dagger - \frac{\kappa_0}{2}\hat{a}_{1C}^\dagger + \sqrt{\kappa_0}\hat{a}_{1C}^{th\dagger}, \tag{34}$$

$$\dot{\hat{b}} = -\frac{\Gamma}{2}\hat{b} - ig^*\hat{a}_{1CC}^\dagger + \sqrt{\Gamma}\hat{b}_{th}, \tag{35}$$

$$\dot{\hat{b}}^\dagger = -\frac{\Gamma}{2}\hat{b}^\dagger + ig\hat{a}_{1CC} + \sqrt{\Gamma}\hat{b}_{th}^\dagger, \tag{36}$$

$$\dot{\hat{d}} = -i\omega_S\hat{a}_{1C} - i\omega_S\hat{a}_{1CC} - (\gamma'_{MAIN} + i\Delta_{MAIN})\hat{d} + iG_0\hat{X} + \sqrt{2\gamma'_{MAIN}}\hat{d}^{th}, \tag{37}$$

$$\dot{\hat{d}}^\dagger = i\omega_S\hat{a}_{1C}^\dagger + i\omega_S\hat{a}_{1CC}^\dagger - (\gamma'_{MAIN} - i\Delta_{MAIN})\hat{d}^\dagger - iG_0\hat{X} + \sqrt{2\gamma'_{MAIN}}\hat{d}^{th\dagger}, \tag{38}$$

$$\dot{\hat{X}} = \hat{P}/M, \tag{39}$$

$$\dot{\hat{P}} = \hbar G_0(\hat{d} + \hat{d}^\dagger) - F_{GW} \tag{40}$$

where $\gamma'_{MAIN} = \gamma_{MAIN} + \gamma_{ADD}$. We define the state vector as

$$\boldsymbol{x}' = (\hat{a}_{1CC}, \hat{a}_{1CC}^\dagger, \hat{a}_{1C}, \hat{a}_{1C}^\dagger, \hat{b}, \hat{b}^\dagger, \hat{d}, \hat{d}^\dagger, \hat{X}, \hat{P})^T, \tag{41}$$

and then we subtract the steady state value to get

$$\boldsymbol{x} = \boldsymbol{x}' - \boldsymbol{x}_{SS}. \tag{42}$$

We then have the state equation for $\boldsymbol{x}$:

$$\dot{\boldsymbol{x}} = A\boldsymbol{x} \tag{43}$$

where A is the state matrix,

$$A \equiv \begin{pmatrix}
-\kappa_0/2 & 0 & 0 & 0 & 0 & -ig^* & -i\omega_S & 0 & 0 & 0 \\
0 & -\kappa_0/2 & 0 & 0 & ig & 0 & 0 & i\omega_S & 0 & 0 \\
0 & 0 & -\kappa_0/2 & 0 & 0 & 0 & -i\omega_S & 0 & 0 & 0 \\
0 & 0 & 0 & -\kappa_0/2 & 0 & 0 & 0 & i\omega_S & 0 & 0 \\
0 & -ig^* & 0 & 0 & -\Gamma/2 & 0 & 0 & 0 & 0 & 0 \\
ig & 0 & 0 & 0 & 0 & -\Gamma/2 & 0 & 0 & 0 & 0 \\
-i\omega_S & 0 & -i\omega_S & 0 & 0 & 0 & -\gamma'_{MAIN} - i\Delta_{MAIN} & 0 & iG_0 & 0 \\
0 & i\omega_S & 0 & i\omega_S & 0 & 0 & 0 & -\gamma'_{MAIN} + i\Delta_{MAIN} & -iG_0 & 0 \\
0 & 0 & 0 & 0 & 0 & 0 & 0 & 0 & 0 & 1/M_{EFF} \\
0 & 0 & 0 & 0 & 0 & 0 & \hbar G_0 & \hbar G_0 & 0 & 0
\end{pmatrix}, \tag{44}$$

The open-loop output of the system is

$$y = C\boldsymbol{x}, \tag{45}$$

where $C$ is a $1\times 10$ row vector. We choose $C = (0,0,0,0,0,0,1/\sqrt{2},1/\sqrt{2},0,0)$, which in turn implies that $y = (\hat{d} + \hat{d}^\dagger)/\sqrt{2}$. The matrix $A$ has some eigenvalues in the upper complex plane, which means the system is unstable. In order to stabilize the system, we use an observer based feedback control [Chapter 9 in Ref. 30] process, which is discussed in more detail in Appendix A. This is illustrated schematically in Fig. 8, where an estimate of the state $\tilde{x}$ is fed back via the control input $u = F\tilde{x}$ in the form:

$$\dot{x} = Ax + Bu. \qquad (46)$$

Here $B$ is a $10 \times 1$ column vector, and $F$ is a $1 \times 10$ row vector representing the feedback gain.

Since the state of the system $x(t)$ cannot be measured directly, we use an estimate $\tilde{x}(t)$ of the state $x(t)$ determined by a state observer [30]. The estimate is constructed as [same as Eq. (A8) in Appendix A]

$$\dot{\tilde{x}} = A\tilde{x} + Bu + K(y - \tilde{y}) = (A - KC)\tilde{x} + Bu + Ky, \qquad (47)$$

where $K$ is the estimate gain. As shown in Fig. 8, for constructing the state estimate $\tilde{x}(t)$, we need the output of an integrator block, which has three inputs: $q(t) = (A - KC)\tilde{x}(t)$, which is the estimator multiplied by the matrix $(A - KC)$, containing the information of the operating parameters of the system; $w(t) = Bu(t)$, which is the input $u$ multiplied by the input matrix $B$; $z(t) = Ky(t)$, which is the output $y$ multiplied by the estimator gain $K$.

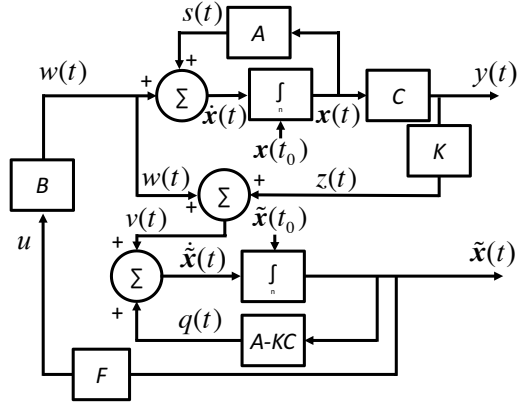

FIG. 8. Schematic illustration of the observer-based controller.

We choose $B$ to be $B = (0,0,0,0,0,0,0,0,1,0)^T$, which corresponds to a situation where the feedback signal is applied to the test mass mirrors ($M_2$) of the main interferometer. In this case, the system is controllable since the controllability matrix [Section 5.2.1 in Ref. 30] $\mathcal{Q} \equiv [B, AB, A^2B, A^3B, A^4B, A^5B, A^6B, A^7B, A^8B, A^9B]$ has full rank of 10, and, therefore, has a non-zero determinant. The same is true for the observability matrix [Section 5.2.2 in Ref. 30] $\mathcal{O} \equiv [C; CA; CA^2; CA^3; CA^4; CA^5; CA^6; CA^7; CA^8; CA^9]$ with the output matrix chosen as

$C = (0,0,0,0,0,0,1/\sqrt{2},1/\sqrt{2},0,0)$.

The parameters such as the decay rate of the phonon and photon modes, the circulating power of the laser, and the waveguide-resonator coupling rate need to be determined first. The corresponding numbers are then encoded in an electronic circuit to realize the matrix $(A - KC)$, and the estimate is multiplied electronically by this matrix to obtain the first input of the integrator, $q(t)$. The second input, $w(t)$, of the integrator is the same as the signal fed back to the original system. The final output of the interferometer, namely the signal produced by the balanced homodyne detector [31], can be expressed as $\zeta(\hat{d}+\hat{d}^{\dagger})/\sqrt{2} = \zeta y$, where $\zeta$ is a known proportionality constant. Thus, the value of y is determined by dividing this signal by $\zeta$, and multiplied electronically by the gain factor $K$ to produce $z(t)$, the third input to the integrator. The choice of initial condition, $\tilde{x}(t_0)$, for the integrator does not affect the behavior of the estimate, since the error between the state and the estimate decays to zero for suitably chosen gain $K$, as explained in detail in Appendix A. As such, a convenient choice for the initial condition is simply $\tilde{x}(t_0) = 0$. The output of the electronic integrator gives the estimate of the system. The electronic output of the estimate is then multiplied by the feedback gain $F$ to get $u$ and then applied to the end mirrors of the interferometer as a feedback signal.

The values for $K$ and $F$ can be chosen so that the eigenvalues of $A$-$KC$ and $A$+$BF$, respectively, are all in the lower half of the complex plane. Therefore, the values also depend on the operating parameters of the system. We choose

$$F = (-2.9\times10^{-3}+6.8\times10^{-5}i, -2.9\times10^{-3}-6.8\times10^{-5}i, 83.0+1.5\times10^{4}i, 83.0-1.5\times10^{4}i, 4.8\times10^{-1}+3.9\times10^{-3}i, \\ 4.8\times10^{-1}-3.9\times10^{-3}i, -2.7\times10^{-1}-2.9\times10^{-7}i, -2.7\times10^{-1}+2.9\times10^{-7}i, -1.0\times10^{7}, -4.9\times10^{21}) \quad (48)$$

$$K = (4.0\times10^{12}+2.8\times10^{14}i, 4.0\times10^{12}-2.8\times10^{14}i, -6.3\times10^{25}+1.8\times10^{23}i, -6.3\times10^{25}-1.8\times10^{23}i, \\ 2.1\times10^{14}+1.1\times10^{21}i, 2.1\times10^{14}-1.1\times10^{21}i, 7.1\times10^{6}-5.8\times10^{20}i, 7.1\times10^{6}+5.8\times10^{20}i, -2.2\times10^{9}, 1.4\times10^{11})^{T} \quad (49)$$

For this choice of parameters, it is possible to achieve the result shown in Fig. 7.

The error signal used for the feedback scheme presented above is proportional to the GW signal. Thus, this feedback scheme can, in principle, be used to replace the current servo for controlling the differential arm lengths, and the GW signal would be extracted from the error signal. In constructing this feedback scheme, we have not taken into account the explicit details of the various complex servos that are already employed in the aLIGO apparatus [16]. It is likely that if/when the system proposed here is implemented, the overall servo system needs to be modified to some extent.

## V. CONCLUSION

We have presented an explicit scheme for realizing a negative dispersion medium (NDM) employing an optomechanical resonator for use in the fast light enhanced compound-cavity signal-recycling (CC-SR) interferometric gravitational wave detector. In contrast to the Rb atom based gain with electromagnetically induced transparency (GEIT) system proposed in Ref. 9, which operates at 795nm, the system proposed here can be realized at 1064nm, which is the wavelength currently used in aLIGO. The NDM makes use of a microresonator supporting optomechanical interaction, with a control field applied at a higher frequency than the probe field. Under proper conditions, the resulting transmission profile of the probe field shows a peak superimposed on an absorption profile, and a negative dispersion. The CC-SR scheme using such an NDM achieves a factor of ~15 enhancement in the quantum-noise-limited sensitivity-bandwidth, where we use the Langevin noise operator model to take into account the noise from the NDM. However, using the parameters required for such sensitivity enhancement, the optomechanical system enters an instability region, where the control field is depleted. We present an observer based feedback control system used to stabilize the CC-SR system.

## ACKNOWLEDGEMENTS

We acknowledge useful discussions with Gaurav Bahl and Seunghwi Kim of University of Illinois, Urbana-Champagne. This work was supported by DARPA through the slow light program under Grant No. FA9550-07-C-0030 and by AFOSR under Grants No. FA9550-10-01-0228 and No. FA9550-09-01-682-0652.

# APPENDIX A: ILLUSTRATION OF THE FEEDBACK CONTROL

We consider the general case where the state equation for the system is [30]

$$\dot{x} = Ax + Bu. \tag{A1}$$

where $x$ is an $n \times 1$ column vector representing the state vector, $A$ is the $n \times n$ state matrix, $B$ is the $n \times m$ input matrix, and $u$ is an $m \times 1$ column vector representing the input signal which will be used eventually for feedback. The output $y$ from the system is a $p \times 1$ column vector:

$$y = Cx. \tag{A2}$$

where $C$ is the $p \times n$ output matrix. For the system considered in the main body, $n = 10$, $m = 1$, and $p = 1$. The block diagram for such a system is shown in Fig. A1(a). The derivative $\dot{x}(t)$ is the sum of two parts: $w(t) = Bu(t)$, which is the control input multiplied by the input matrix; $s(t) = Ax(t)$, which is the state vector multiplied by the state matrix.

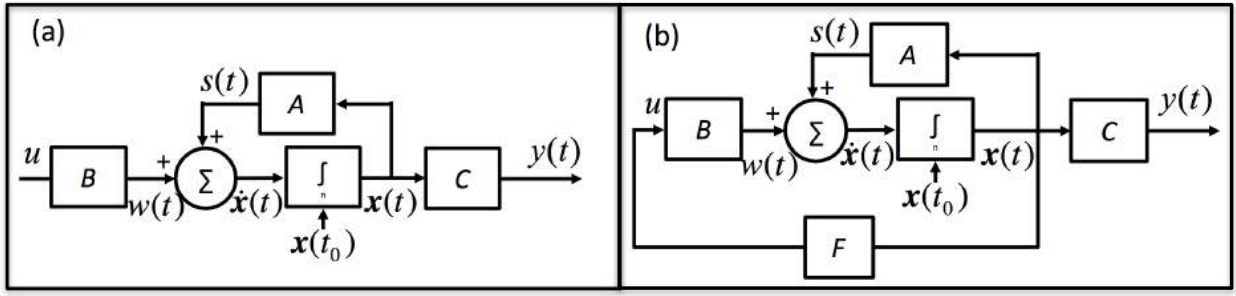

FIG. A1. (a) Block diagram of the open-loop system described in Eqs. (A1) and (A2); (b) Block diagram of the closed-loop system where

$$u = Fx.$$

When the matrix $A$ has eigenvalues in the upper complex plane, the system is unstable. In order to stabilize the system, we need to use a feedback as the control input. If the state vector $x$ can be measured directly, we can feed back the state vector, i.e. $u = Fx$, as shown in Fig. A1(b), where $F$ is an $m \times n$ matrix representing the feedback gain [30]. In this case, Eq. (A2) becomes

$$\dot{x} = (A + BF)x \tag{A3}$$

We can choose a proper gain $F$, so that all the eigenvalues of $A+BF$ are in the lower complex plane. However, in some situation, the state vector cannot be measured directly. In that case, we need to create an estimate of the state vector, $\tilde{x}$, which is an $n\times 1$ column vector. With the knowledge of the matrices A and B and the input u, we can construct the state estimate electronically as follows [Fig. A2(a)]:

$$\dot{\tilde{x}} = A\tilde{x} + Bu. \tag{A4}$$

The state estimate $\tilde{x}(t)$ is the output of an integrator block, whose input is the sum of two parts: the control input multiplied by the input matrix, $w(t) = Bu(t)$; and the state estimate multiplied by the state matrix, $\tilde{s}(t) = A\tilde{x}(t)$. If the estimate is exactly the same as the state, i.e. $\tilde{x}(t) = x(t)$, we have $\tilde{s}(t) = s(t)$. Therefore, the two input parts for $\dot{\tilde{x}}(t)$ and $\dot{x}(t)$ are the same. In Fig. A2(b), we show the closed-loop system where the estimate $\tilde{x}(t)$ is fed back as the control input $u(t) = F\tilde{x}(t)$. In this case, we get the closed-loop equation, i.e., Eq. (A3), since $\tilde{x}(t) = x(t)$.

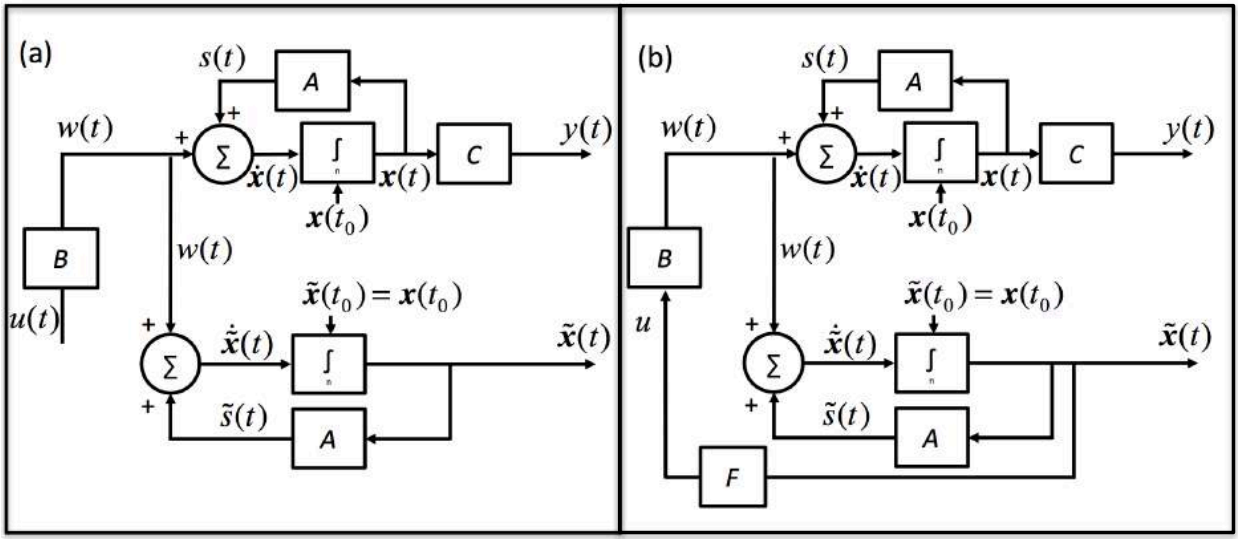

FIG. A2. Block diagram of (a) the open-loop system and (b) the closed-loop system using the estimate in Eq. (A4).

We now discuss the validity of the assumption that $\tilde{x}(t) = x(t)$. From Eqs. (A1) and (A4), the error of the estimate, defined as $e(t) = x(t) - \tilde{x}(t)$, evolves as

$$\dot{e} = Ae. \tag{A5}$$

If we know what the state is at an initial time [i.e. $x(t_0)$], then we can use the same initial condition for the estimate, $\tilde{x}(t_0) = x(t_0)$. In this case, we have that $e(t_0) = 0$, and the error stays zero, i.e. $e(t) = 0$, so that $x(t) = \tilde{x}(t)$. On the other hand, if $e(t_0) \neq 0$, the error will diverge, since $A$ has eigenvalues in the upper complex plane.

In the case where $x(t_0)$ is unknown and we have $e(t_0) \neq 0$, in order to make the error decay to zero, we need to modify the estimate by including a term depending on the error, in a manner so that the equations for the error becomes:

$$\dot{e} = (A - U)e, \tag{A6}$$

and choose a proper $n \times n$ matrix $U$ so that $(A-U)$ has all the eigenvalues in the lower complex plane. This can be achieved by incorporating a term, depending on the error, in the state estimate:

$$\dot{\tilde{x}} = A\tilde{x} + Bu + U(x - \tilde{x}), \tag{A7}$$

which means that we need an extra input part for $\dot{\tilde{x}}(t)$. However, since $x(t)$ is not directly measurable, it is not possible to create the extra input term, $U(x - \tilde{x})$, directly. Instead, we make use of the fact that the output is proportional (via a matrix) to $x$: $y(t) = Cx(t)$. Thus, if we construct an estimate for the output with the same matrix proportionality, namely, $\tilde{y}(t) = C\tilde{x}(t)$, then we get $y - \tilde{y} = C(x - \tilde{x})$. To match dimensionality, we need to multiply this by another $n \times p$ matrix $K$. Thus, $K(y - \tilde{y}) = KC(x - \tilde{x}) = U(x - \tilde{x})$, and the equations for the state estimate now becomes:

$$\dot{\tilde{x}} = A\tilde{x} + Bu + K(y - \tilde{y}) = (A - KC)\tilde{x} + Bu + Ky. \tag{A8}$$

This is equivalent to Eq. (A7) when $U = KC$, which means that the error of the estimate $e(t)$ follows Eq. (A6). In this case, with a suitably chosen gain $K$, the value we use for the initial condition of the estimate would not affect the behavior of the estimator since the error $e(t)$ will always decay to zero. The resulting block diagram is shown in Fig. A3(a). The state estimate $\tilde{x}(t)$ is the output of an integrator block, whose input is the sum of three parts: $q(t) = (A - KC)\tilde{x}(t)$, which is the estimator multiplied by the matrix $(A - KC)$; $w(t) = Bu(t)$, which is the input $u$ multiplied by the input matrix $B$; $z(t) = Ky(t)$, which is the output $y$ multiplied by the estimator gain $K$. The initial condition is chosen as $\tilde{x}(t_0) = 0$. When $\tilde{x}(t) = x(t)$, the term $KC\tilde{x}(t)$ in $q(t)$ cancels out $z(t)$, and we get back to the form in Eq. (A4) [shown in Fig. A2(a)]. The estimate can be fed back to the input $u(t) = F\tilde{x}(t)$ as shown in Fig. A3(b).

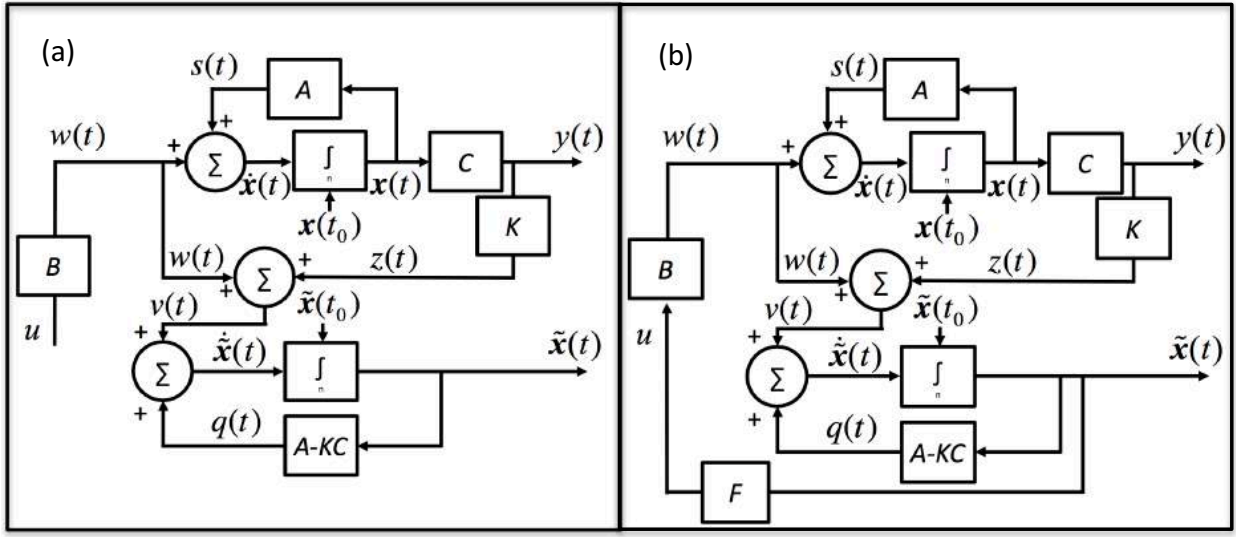

FIG. A3. Block diagram of (a) the open-loop system and (b) the closed-loop system using the estimate described in Eq. (A8).

To analyze the behavior of this closed-loop system, we plug $u(t) = F\tilde{x}(t)$ into Eqs. (A4) and (A8), and get the state equations:

$$\dot{x} = Ax + BF\tilde{x}, \tag{A9}$$

$$\dot{\tilde{x}} = KCx + (A - KC + BF)\tilde{x}, \tag{A10}$$

from which, we get the evolution of the estimate error as:

$$\dot{e} = (A - KC)e, \tag{A11}$$

We can rewrite Eq. (A9) and Eq. (A11) in the matrix form:

$$\begin{pmatrix} \dot{x} \\ \dot{e} \end{pmatrix} = \begin{pmatrix} A+BF & -BF \\ 0 & A-KC \end{pmatrix} \begin{pmatrix} x \\ e \end{pmatrix}, \tag{A12}$$

To stabilize the closed-loop system, we need to place all the eigenvalues of $A$-$KC$ and $A$+$BF$ in the lower half of the complex plane, which can be assigned via $K$ and $F$ provided that the system considered in Eqs. (A1) and (A2) is controllable [Section 5.2.1 in Ref. 30] and observable [Section 5.2.2 in Ref. 30]. The pair ($A$, $B$) is controllable when the controllability matrix [30] $\mathcal{Q} \equiv [B, AB, A^2B, ..., A^{n-1}B]$ has full rank of $n$, and, therefore, has a non-zero determinant. The pair ($A$, $C$) is observable when the observability matrix [30] $\mathcal{O} \equiv [C; CA; CA^2; ...; CA^{n-1}]$ has full rank of $n$.

# APPENDIX B: BEAM SPITTER/COMBINER USING TRIANGULAR CAVITY

The beam splitter/combiner (G1 and G2) used in Fig. 5 requires a high resolution, and can be realized using a triangular optical cavity as shown in Fig. A4(a).

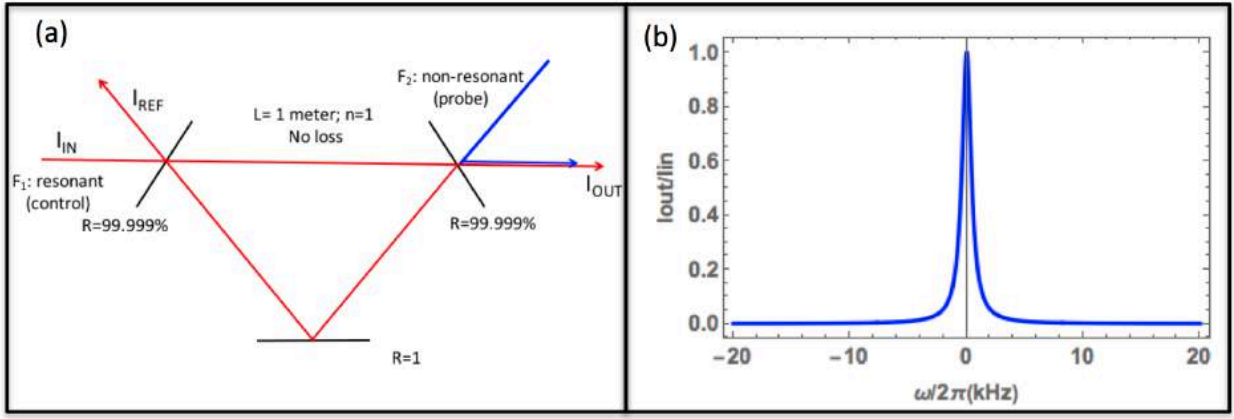

FIG. A4. (a) Schematic illustration of the triangular optical cavity as the beam combiner; (b) Plot of the transmissivity $I_{out}/I_{in}$ as a function of the frequency $\omega/(2\pi)$.

For the input beam $F_1$, the transmissivity $I_{OUT}/I_{IN}$ and reflectivity $I_{REF}/I_{IN}$ are

$$\frac{I_{OUT}}{I_{IN}} = \left[1 + \frac{4R}{T^2}\sin^2(\omega/FSR)\right]^{-1} \tag{A12}$$

$$\frac{I_{OUT}}{I_{IN}} = 1 - \left[1 + \frac{4R}{T^2}\sin^2(\omega/FSR)\right]^{-1} \tag{A13}$$

where the free spectrum range is $FSR = c/(nL)$, $n=1$ is index of refraction of the medium inside the triangular cavity, $R$ and $T=1-R$ are the reflectivity and transmissivity of the two mirrors in the cavity, and the third mirror is assumed to be perfectly reflective. The full width half maximum and the finesse of the triangular cavity are then

$$FWHM = \frac{2c}{\pi nL}\arcsin\left(\frac{T}{2\sqrt{R}}\right), \tag{A14}$$

If the mirror reflectivity is taken to be R=99.999%, FWHM is as small as ~0.9kHz and the FSR is 300MHz. The transmissivity of the cavity in this case is plotted in Fig. A4(b). Such a cavity can therefore be used to separate or combine beams with a frequency different as low as 1MHz.

For the triangular cavity to function as a beam combiner, the control laser (indicated as $F_1$) is resonant in the cavity while the probe (indicated as $F_2$) is non-resonant in the cavity. On the other hand, for the triangular cavity to function as a beam splitter, both beams are sent in the opposite directions. The part of the signal recycling cavity incorporating the triangular optical cavities as G1/G2 are shown in Fig. A5. A small amount of control field may be reflected by the triangular cavity and contribute to the output of the interferometer. However, it should be noted that the detection process employs mixing with the main pump laser. Since the control field frequency differs from the main pump laser by 1GHz, the beat note between the leaked control field and the main pump laser can be easily filtered out electronically, without any effect on the GW signal. Also, in aLIGO, the signal is sent through an output mode cleaner (OMC) before detection [16]. The OMC would filter out such leaked control light as well.

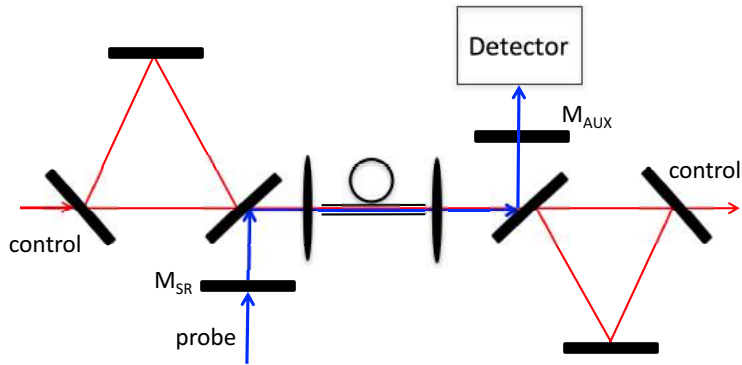

FIG. A5. Schematic illustration of the signal recycling cavity incorporating the triangular optical cavities as G1/G2.

**References**

1. M. Salit and M.S. Shahriar, "Enhancement of Sensitivity-Bandwidth Product of Interferometric Gravitational Wave Detectors using White Light Cavities", J. Opt. 12, 104014 (2010).

2. R. H. Rinkleff and A. Wicht, "The concept of white light cavities using atomic phase coherence", Phys. Scr. T118 85-88 (2005).

3. A. Wicht, R. H. Rinkleff, L. S. Molella and K. Danzmann, "Comparative study of anomalous dispersive transparent media", Phys. Rev. A 66 063815 (2002).

4. A. Rocco, A. Wicht, R. H. Rinkleff and K. Danzmann, "Anomalous dispersion of transparent atomic two- and three-level ensembles", Phys. Rev. A 66 053804 (2002).

5. A. Wicht, M. Muller, R. H. Rinkleff, A. Rocco and K. Danzmann, "Experimental demonstration of negative dispersion without absorption", Opt.Commun. 179 107–15 (2000).

6. G. S. Pati, M. Salit, K. Salit, and M. S. Shahriar, "Demonstration of a tunable-bandwidth white light interferometer using anomalous dispersion in atomic vapor", Phys. Rev. Lett. 99, 133601 (2007).

7. H. N. Yum, M. Salit, G. S. Pati, S. Tseng, P. R. Hemmer, and M. S. Shahriar, "Fast-Light in a Photorefractive Crystal for Gravitational Wave Detection", Opt. Express 16, 20448-20456 (2008).

8. M. Zhou, Z. Zhou, and S. M. Shahriar, "Quantum noise limits in white-light-cavity-enhanced gravitational wave detectors", Phys. Rev. D. 92, 082002 (2015).

9. M. Zhou, Z. Zhou, and S. M. Shahriar, "Realization of Gain with Electromagnetically Induced Transparency System with Non-degenerate Zeeman Sublevels in $^{87}$Rb", arXiv:1609.09838.

10. Bahl, G., Zehnpfennig, J., Tomes, M., and Carmon, T. "Stimulated optomechanical excitation of surface acoustic waves in a microdevice", Nature Commun. 2, 403 (2011).

11. Bahl, G., Tomes, M., Marquardt, F. & Carmon, T. "Observation of spontaneous Brillouin cooling", Nature Phys. 8, 203–207 (2012).

12. J. Kim, M. C. Kuzyk, K. Han, H. Wang, and G. Bahl, "Non-reciprocal Brillouin scattering induced transparency", Nature Phys. 11, 275 (2015).

13. M. Aspelmeyer, T. J. Kippenberg, and F. Marquardt, "Cavity optomechanics", Rev. Mod. Phys. 86, 1391 (2014).

14. It is well know that for a causal system, the transmission and dispersion profiles for a probe are governed by the Kramer-Kronig relations. For most systems, there is  a general correspondence between the shape of the dispersion and the shape of the transmission. Specifically, a peak (dip) in the transmission profile corresponds to a positive (negative) dispersion. However, this correspondence is not universal. In the case of light coupled from waveguides to resonators, it is sometimes the case that a peak (dip) in the transmission profile corresponds to a negative (positive) dispersion, without violating the Kramers-Kronig relations. The case considered here is one example of such counter-conventional correspondence. We have also found other cases of such counter-conventional correspondence.

15. M. O. Scully and M. S. Zubairy, *Quantum optics* (Cambridge University Press, 1997), Chap. 12.

16. J. Aasi et al., "Advanced LIGO," Class. Quantum Grav. 32, 074001 (2015).

17. C. M. Caves and B. L. Schumaker, "New formalism for two-photon quantum optics. I. Quadrature phases and squeezed states," Phys. Rev. A **31**, 3068 (1985).

18. B. L. Schumaker and C. M. Caves, "New formalism for two-photon quantum optics. II. Mathematical foundation and compact notation," Phys. Rev. A **31**, 3093 (1985).

19. A. Buonanno and Y. Chen, "Quantum noise in second generation, signal-recycled laser interferometric gravitational-wave detectors," Phys. Rev. D 64,042006 (2001).

20. G. Rempe, R. J. Thompson, and H. J. Kimble, "Measurement of ultralow losses in an optical interferometer," Optics Letters, Vol. 17, No. 5 (1992).

21. J. Hiller, J. D. Mendelsohn and M. F. Rubner, "Reversibly erasable nano-porous anti-reflection coatings from polyelectrolyte multilayers," Nature Materials, Vol. 1, 59-63 (2002).

22. R.A.Norte, J. P. Moura, and S. Gröblacher, "Mechanical Resonators for Quantum Optomechanics Experiments at Room Temperature," Phys. Rev. Lett. **116**, 147202 (2016).

23. Tina Müller, Christoph Reinhardt, and Jack C. Sankey, "Enhanced optomechanical levitation of minimally supported dielectrics," Phys. Rev. A **91**, 053849 (2015).

24. M. L. Gorodetsky, A. A. Savchenkov, and V. S. Ilchenko, "Ultimate Q of optical microsphere resonators," Opt. Lett. 21, 453-455 (1996).

25. D. W. Vernooy, V. S. Ilchenko, H. Mabuchi, E. W. Streed, and H. J. Kimble, "High-Q measurements of fused-silica microspheres in the near infrared," Opt. Lett. 23, 247-249 (1998).

26. H. Miao, Y. Ma, C. Zhao, and Y. Chen, "Enhancing the Bandwidth of Gravitational-Wave Detectors with Unstable Optomechanical Filters," Phys. Rev. Lett. **115**, 211104 (2015).

27. A. Buonanno and Y. Chen, "Signal recycled laser-interferometer gravitational-wave detectors as optical springs," Phys. Rev. D. **65**, 042001 (2002).

28. A. Buonanno and Y. Chen, "Scaling law in signal recycled laser-interferometer gravitational-wave detectors," Phys. Rev. D. **67**, 062002 (2003).

29. A. M. Jayich, J. C. Sankey, B. M. Zwickl, C. Yang, J. D. Thompson, S. M. Girvin, A. A. Clerk, F. Marquardt, and J. G. E. Harris, "Dispersive optomechanics: a membrane inside a cavity," New Journal of Physics **10**, 095008 (2008).

30. P. J. Antsaklis and A. N. Michel, *A Linear Systems Primer* (Birkhauser, Boston, 2007).

31. M S Stefszky, C M Mow-Lowry, S S Y Chua, D A Shaddock, B C Buchler, H Vahlbruch, A Khalaidovski, R Schnabel, P K Lam and D E McClelland, "Balanced homodyne detection of optical quantum states at audio-band frequencies and below," Class. Quantum Grav. 29, 145015 (2012).